*Title* Evolutionary drivers, morphological evolution and diversity dynamics of a surviving mammal clade: cainotherioids at the Eocene-Oligocene transition

*Authors*

Weppe, R[1], Orliac, MJ[1], Guinot, G[1], Condamine, FL[1]

*Affiliations*

[1]Institut des Sciences de l'Évolution de Montpellier, Université de Montpellier, CNRS, IRD, EPHE, Place Eugène Bataillon, 34095 Montpellier Cedex 5, France.

**Corresponding author**

romain.weppe@umontpellier.fr

*Abstract*

The Eocene-Oligocene transition (EOT) represents a period of global environmental changes particularly marked in Europe and coincides with a dramatic biotic turnover. Here, using an exceptional fossil preservation, we document and analyse the diversity dynamics of a mammal clade, Cainotherioidea (Artiodactyla), that survived the EOT and radiated rapidly immediately after. We infer their diversification history from Quercy Konzentrat-Lagerstätte (South-West France) at the species level using Bayesian birth-death models. We show that cainotherioid diversity fluctuated through time, with extinction events at the EOT and in the late Oligocene, and a major speciation burst in the early Oligocene. The latter is in line with our finding that cainotherioids had a high morphological adaptability following environmental changes throughout the EOT, which likely played a key role in the survival and evolutionary success of this clade in the aftermath. Speciation is positively associated with temperature and continental fragmentation in a time-continuous way, while extinction seems to synchronize with environmental change in a punctuated way. Within-clade interactions negatively affected the cainotherioid diversification, while inter-clade competition might explain their final decline during the late Oligocene. Our results provide a detailed dynamic picture of the evolutionary history of a mammal clade in a context of global change.

# 1. Introduction

The evolution of biological diversity is determined by the interaction between speciation and extinction rates, which lead clades to thrive or decline [1–4]. Speciation and extinction are in turn influenced by biotic and abiotic factors that vary over time and space, and understanding the role of these factors represents one of the main goals of evolutionary biology [3,5–8]. The evolutionary history of clades has been demonstrated to be influenced by environmental changes such as temperature variations or sea-level fluctuations [9–13], and/or biotic interactions such as predation and competition for resources [14–17].

In the Cenozoic, the Eocene-Oligocene transition (EOT) marks a period of major environmental changes worldwide. The EOT is associated with a global cooling [18,19], the onset of Antarctic glaciation [20,21], a major eustatic sea-level fall [22,23], and concurrent disruption of marine and continental ecosystems [24]. Among the latter, a major faunal turnover, known as the 'Grande Coupure' [25], has been documented in Europe. This event corresponds to a dramatic extinction phase of European endemic placental mammals [26,27], which evolved in an insular context during the middle and late Eocene [26,28], coupled with the arrival of immigrant taxa from Asia [26,28–31].

Knowledge of the European mammalian turnover during the EOT is widely based on the abundant material collected from the karstic infillings of the Quercy phosphorites (South-Western France) [27–29,32]. This fossiliferous area covers over 200 localities documenting regional faunal assemblages with a 1-million-year resolution [33,34], and a continuous record covering the late Eocene to late Oligocene interval (38 to 25 million years ago, Mya). The abundance of fossil material retrieved in Quercy makes it a multi-site Konzentrat-Lagerstätte, and the last 50 years of excavations have built an unprecedented collection providing a unique opportunity to study the processes and mechanisms of diversification of mammalian clades.

Artiodactyls are a major component of the European endemic placental mammals

[28,29]. They constitute one the most diverse mammalian groups during the Eocene, but are drastically impacted by the Grande Coupure with numerous artiodactyl families disappearing around the EOT [28]. However, few artiodactyl groups persisted after the Grande Coupure, the Cainotherioidea being one of them. This small rabbit-sized taxon, documented from the late Eocene to the middle Miocene in the fossil record of Western Europe [35–37], was particularly flourishing during the Oligocene [32,37], and represents the most abundant artiodactyl clade retrieved in the Quercy fossil record. Consequently, Cainotherioidea represents a relevant case study to investigate the diversity dynamics of a survivor clade in a pre- and post-crisis context based on an unprecedented preservation rate.

In this work, we have revised, compiled and analysed the fossil record of Cainotherioidea (more than 900 occurrences) from 37 late Eocene to late Oligocene Quercy localities. We further tested whether and to which extent climatic and geographic changes, biotic interactions, as well as the evolution of life history (body mass) and diet (dental and mandibular morphology) traits, affected the speciation and extinction rates of this clade. The focus on this exceptional fossil record concentrated to a restricted geographical area substantially reduces the number of confounding effects (e.g. preservation bias and differences in climate conditions and biologic interactions [5,38–40]), and provides access to a detailed, dynamic picture of the evolutionary history of a clade of terrestrial mammals in a general context of global change.

## 2. Material and Methods

### (a) Dataset of Cainotherioidea Species

We compiled all species-level fossil occurrences of Cainotherioidea from 37 localities of the Quercy area. The material is housed in the collections of the University of Montpellier (France). Species occurrences at each locality result from direct revision and determination of the material (dental, mandibular, and cranial). Each Quercy locality was associated with both a Mammalian Paleogene (MP) reference level [33,41] and a numeric age (electronic supplementary material, **tables S1-S2**). The resulting dataset comprises 915 occurrences for 15 species (electronic supplementary material, **dataset S1**), ranging from the late Eocene to the late Oligocene (38 to 25 Mya).

We investigated six cranio-dental measurements (electronic supplementary material, **figure S1 and dataset S2**) to test for trait correlation between diversification and environment: (i) width of the P/3 (LP3); (ii) diastemata length (DIAS); (iii) mandibular corpus height (HCM3); (iv) mandibular symphysis height (HSM); (v) ramus angulus (ANGLE); (vi) lower molar row length (LMRL). The first five traits relate to diet [32,42–45], whereas LMRL is a proxy for body mass [46]. LP3 and DIAS are indicators of the chewing surface [32], HCM3 and HSM of the gracility/robustness of the mandibular corpus [42,44], and ANGLE of the chewing movement degree [47]. These variables are standardized by the body mass proxy (LMRL).

**(b) Selection of Abiotic Variables**

We focused on the role of three abiotic variables (temperature, sea level, continental fragmentation; see electronic supplementary material, **dataset S3**) which have been shown to impact past mammal diversifications [10,13,15,48–51] and spanning the full-time range of Cainotherioidea. Major trends in global climate change through time are estimated from relative proportions of different oxygen isotopes ($\delta^{18}O$) in samples of benthic foraminiferal shells [52]. Trends in global temperature changes over time were obtained from $\delta^{18}O$ data transformed into deep-sea temperature estimates [53]. Trends in global sea-level changes over time were obtained from $\delta^{18}O$ data indirectly recorded in the foraminifera shell chemistry [54]. The continental fragmentation is approximated with an index obtained using paleogeographic reconstructions at 1-million-year time intervals [55].

**(c) Diversity and Correlation Analyses**

We analysed the fossil dataset with PyRate [56] under the birth-death model with constrained shifts (BDCS; [57]) to simultaneously estimate the preservation process ($q$), the times of speciation ($Ts$) and extinction ($Te$) of each species, the speciation ($\lambda$) and extinction ($\mu$) rates, and their variations through time. We ran PyRate for 10 million Markov chain Monte Carlo (MCMC) generations with time bins of one million year and two million years, and sampling every 1,000 generations to approximate the posterior distribution of all parameters. Given the excellent quality of the considered fossil record, all analyses were performed with a homogeneous Poisson process of preservation and accounted for varying preservation rates across taxa using the Gamma model [58]. We replicated the analyses 10 times by randomizing fossil ages under an uniform law, and we combined the posterior estimates across all replicates to generate rates-through-time plots (speciation, extinction, and net diversification) and past diversity dynamics.

We used the Covar birth–death model [58] to test for the influence of six continuous cranio-dental characters (see above) on the diversification dynamics of cainotherioids over time. We also performed statistical treatment of trait measurements using R 3.6.2 [59], by combining the fossil occurrences of the first dataset, their dating (fossil ages from PyRate replicate; MP level locality dating) and the intraspecific trait variation (see electronic supplementary material**, appendix S1 and dataset S4**). We compared trait values through time by plotting median and variance of the focal traits by MP level. We used a non-parametric Wilcoxon's test to test for difference (at a 5% risk) in the trait values during, before and after the EOT. Difference (at a 5% risk) in the trait dispersion during, before and after the EOT, was tested using a non-parametric Siegel-Tukey's test (PMCMRplus package

[60]). Correlation (at a 5% risk) between traits and abiotic variables was tested using non-parametric Kendall's tau and Spearman's rho tests. We assigned to each fossil occurrence a value from each of the three abiotic variables according to their corresponding age as estimated with PyRate (see electronic supplementary material, **appendix S2**).

We used the multivariate birth-death (MBD) model of PyRate [61] to assess whether environmental variables explain temporal variations in speciation and extinction rates of cainotherioids. Under the MBD model, $\lambda$ and $\mu$ can change through correlations with time-continuous variables and the strength and sign of the correlations are jointly estimated for each variable. PyRate jointly estimates the baseline speciation ($\lambda_0$) and extinction ($\mu_0$) rates and all correlation parameters ($G_\lambda$ and $G_\mu$) using a horseshoe prior to control for over-parameterization and for the potential effects of multiple testing [61]. We ran the MBD model using 10 million MCMC iterations and sampling every 1,000 generations to approximate the posterior distribution of all parameters ($\lambda_0$, $\mu_0$, 4 $G_\lambda$, 4 $G_\mu$), and the shrinkage weights ($\omega$) of each correlation parameter.

Finally, we estimated the probability for a lineage to become extinct as a function of its age (the elapsed time since its origination) by fitting the age-dependent extinction (ADE) model [62]. We ran PyRate for 10 million MCMC generations with a time-variable Poisson process of preservation, while accounting for varying preservation rates across taxa using the Gamma model. We replicated the analyses on 10 randomized datasets and combined the posterior estimates across all replicates. Scripts are provided in electronic supplementary material, **appendix S2-S3**.

## Results and Discussion

**Diversity Dynamics of a Surviving Clade.** The EOT was a time of massive turnover and is now considered as a major extinction event [24,26]. A number of groups disappeared at the end of the Eocene with no relatives in the Oligocene, especially among mammals [10,26,27,29]. Here we document the diversity dynamics of a mammal clade that survived the EOT and radiated dramatically immediately after. We inferred the diversification history of cainotherioids based on 915 occurrences at the species level, representing 15 species. Using the BDCS model, we estimated a preservation rate for cainotherioids at ~14 occurrences/taxon/Myrs, underlining the exceptional quality of the karstic Quercy fossil record for that clade (electronic supplementary material, **table S3**). This rate is comparable to those recovered for invertebrate marine organisms [63–65] and, to our knowledge, is the highest rate reached for a terrestrial vertebrate clade to date (North America Cenozoic: ~6.95 occurrences/taxon/Myrs for Caninae, ~8.04 for Ursidae [15]; Cenozoic: ~1-3 for Rhinocerotidae [58], ~0.25-1 for ruminants [66]; Mesozoic and Cenozoic: ~1.21 for amphibians [67]; ~1.21 for crocodiles, ~0.79 for turtles, and ~1.31 for lepidosaurs [68]). The high preservation rate for cainotherioids is consistent with the highly-sampled and high-quality fossiliferous record of the Quercy Konzentrat-Lagerstätte [29,32,33].

Our results indicate, for time bins of one (**figure 1**) and two million years (electronic supplementary material, **figure S2**), that the diversification dynamics of cainotherioids conforms to a time-variable birth–death process with early high rates of speciation that decreased toward the EOT, increased drastically just after the EOT, and decreased again in the last 5 Myrs (**figure 1A**). Extinction was fairly constant but peaked at the EOT and in the last 5 Myrs (**figure 1B**). This resulted in a positive net diversification early in the clade's history, followed by a 2 Myrs period of negative net diversification throughout the EOT, and followed by a high net diversification in the early Oligocene, and a decline phase in the last 5 Myrs (**figure 1C**). As a consequence, the diversity trajectory of cainotherioids shows a complex evolution over time (**figure 1D**), with an alternation of three successive peaks of species diversity. The first one, recorded in the late Eocene, corresponds to the differentiation of the Robiacinidae; the second one, observed at the very end of the Eocene, corresponds to the occurrence of the first Cainotheriidae species; and the third and major one, occurring in the early Oligocene, reflects the diversification of the subfamily Cainotheriinae. This early Oligocene diversification event (**figures 1A, 1D**), already mentioned in the literature [32,37], shows the successful adaptive response of this clade to the new post-EOT environmental conditions. However, cainotherioids clearly experience an extinction event at the EOT (**figures 1B, 1D**), implying that despite their relative evolutionary success after the transition, this survivor clade was nevertheless significantly impacted by the environmental changes. In addition, speciation rates decrease drastically at the end of the early Oligocene and remain particularly low until the late Oligocene (**figure 1A**), failing to compensate for the successive extinction peaks. Our results thus highlight the importance of the speciation decrease in the decline of clades' diversity that, under a general regime of high background extinction rates, drive clades to extinction [2,4]. These observations also stand for time bins of two million years (electronic supplementary material, **figure S2**).

**Evolutionary Drivers of the Cainotherioid Diversity.** By fitting the ADE model to species occurrences, we found no evidence for ADE in cainotherioids, meaning that young and old species are equally likely to become extinct (electronic supplementary material, **table S4**). These results are in line with the law of constant extinction as envisioned in the Red Queen hypothesis [69] and differ from previous studies on ADE, which mostly found an inverse relationship between taxon age and extinction risk [17,62,70]. Interestingly, the mean longevity of cainotherioid species is relatively low (~3.7 Myrs; electronic supplementary material, **table S4**) as compared to some estimates of mean duration of artiodactyl species: ~5.6 Myrs in the Neogene of Old World [71] or ~4.4 Myrs in the Cenozoic of North America [72]. This could be explained by the limited geographical range of cainotherioid species that were restricted to the Quercy area (~1800 km$^2$), which may have impacted their probability of extinction [73]. Indeed, cosmopolitan taxa seem to have a lower risk of extinction than endemic or regionally restricted species [17,63]. In addition, differences in species longevity can also relate to the time interval of our study that coincides with a period of important climatic change [19,20,53]. A particularly low mean duration is indeed estimated for North American artiodactyl species during the late Eocene-Oligocene period (~2.9 Myrs; [74]), and small longevity values of cainotherioids species might also reflect a general context of environmental crisis. These results, however, alone cannot explain the decline in cainotherioid diversity in Quercy during the late Oligocene.

The impact of environmental variations in speciation and extinction rates of cainotherioids over the considered time span was investigated using the MBD model. We recovered no effects of sea level variations on speciation and extinction rates ($\omega<0.5$; electronic supplementary material**, table. S5**), suggesting that eustatic fluctuations did not influence the diversification of cainotherioids. However, we found evidence for an effect of both continental fragmentation and temperature variations over speciation rates ($\omega>0.5$; electronic supplementary material**, table S5**). These correlations are both positive (G$\lambda_{temperature}$=0.01, G$\lambda_{fragmentation}$=4.61; electronic supplementary material**, table S5**), suggesting higher speciation rates during warmer periods with more fragmented landmasses. The influence of tectonics and global climate change on deep-time biodiversity dynamics has been reported in numerous studies at the mammalian scale, over long-time ranges and broadly distributed groups [10,48–51,75] or over smaller time ranges and/or restricted groups [15,76,77]. Europe experienced important climatic and tectonic fluctuations between the Eocene and Oligocene, with (i) a significant global drop in temperature at the EOT followed by successive phases of warming at the beginning and the end of the Oligocene [53,78]; and (ii) an important phase of Alpine orogeny [31,79], which led to a decrease in continental fragmentation during the EOT [55]. By positively affecting the cainotherioid capacity to diversify, our study highlights that continental fragmentation and temperature are likely drivers of speciation rates. The effect of these two factors on speciation is well known and has been tested in other studies in vertebrates (amphibians: [67]; mammals: [10]; rodents: [76]; sharks: [17]; tetrapods: [51]).

While global environmental change affected the speciation of cainotherioids, we found no significant long-term effects of abiotic factors over the observed trends in extinction rates throughout the Eocene-Oligocene period (electronic supplementary material**, table S5**). However, we observed that peaks of extinction occur almost concomitantly with major environmental changes, notably at the EOT (global cooling and sea-level fall; [19,21,23]) and at the end of the Oligocene (global warming; [52,80]) (**figure 1B**). This highlights the influence of environmental variables on cainotherioid extinction over short time bins, which may be reminiscent of the Court Jester model [5,81]. These punctual extinction events occur in a context of abiotic variations, regardless of their increase or decrease, which might explain the lack of long-term correlation between extinction rates and abiotic factors. This point highlights the limitation of our study with the MBD model in its current version, where it is only possible to infer time-continuous correlations and not punctual correlations.

The decline of caintherioids after their rapid diversification in the early Oligocene (**figure 1D**) partly resulting from their failure to speciate, illustrates one of the most pervasive principles in macroevolution (i.e. diversification is slowing down; [51,82,83]) and is not explained by the abiotic factors considered here. However, the MBD analyses indicate that within-clade diversity-dependent processes played a role on cainotherioid diversification. The results indicate a negative correlation ($G\lambda_{Diversity}=-2.65$) between speciation and clade diversity ($\omega>0.5$; electronic supplementary material**, table S5**), implying that cainotherioid speciation rates decreased as they diversified over time. This negative within-clade diversity-dependence can be linked with ecological constraints caused by competition for resources or niche availability (see below; [82–84]) that probably constrained the diversity of cainotherioid species. Moreover, inter-clade competition may also have contributed to the Oligocene decline of cainotherioids. The arrival of new competitors from Asia during the Grande Coupure [25], like ruminants, glires (Castoridae, Cricetidae and Leporidae) and carnivores [28,29,84], may have exerted significant selective pressure on cainotherioids. The fossil record indicates that these newcomer clades diversified after their arrival in Europe [28,29,66,86–87] and may have competed with cainotherioids already present by sharing similar diets [28,29] or by creating new predator-prey interactions. This part of the competition between clades could not be treated in this study, requiring a compilation and revision of fossil occurrences data similar to the cainotherioids but for three large groups of mammals present in Europe.

**Morphological Evolution of Caintherioid Species after the EOT.** In spite of their general conservative cranio-dental and postcranial morphology throughout their fossil record, Oligocene cainotherioid species seem to differ from Eocene relatives by a generally larger size (body mass) and augmented chewing surface [32]. Consequently, we tested for a correlation between cranio-dental morphology and the diversity dynamics of cainotherioids, and investigated six variables associated to body mass and diet (see above) as drivers of the diversification of the clade. Our results indicate that none of the traits tested have a significant effect on speciation and extinction rates over the time interval considered (electronic supplementary material**, table S6**). However, macroscopic observations indicate that traits do differ between species before and after the EOT: Wilcoxon (difference in the trait) and Siegel-Tukey (dispersion in the trait) tests show significant evidence that traits are distinct between species throughout the transition (electronic supplementary material**, table S7**). Oligocene cainotherioids, mostly the subfamily Cainotheriinae, are more speciose and show an increase in their size range (**figure 2;** electronic supplementary material, **figure S3A**), which occurred shortly after the EOT and could be related to the clade's evolutionary success because it coincides with the early Oligocene diversification phase. At their diversity peak (MP 23, *ca*. 31 Ma), eight species cohabited (**figure 1D**), covering a wide range of body sizes; large and small species then persisted until the end of the Oligocene. After the early Oligocene diversification of Cainotheriinae, members of the group all display a very similar, highly derived, dental morphology [32,37] and most probably did not drastically differ in their diet. These morphological and dietary similarities might thus have triggered within-clade competition and niche saturation among Oligocene taxa as suggested by the MBD model.

According to our results, compared with Eocene ones, Oligocene species display a larger chewing surface (expressed through LP3 and DIAS; **figure 3A;** electronic supplementary material, **figure S3B and table S7**), and more robust mandibles (HCM3 and HSM; **figure 3B;** electronic supplementary material, **figure S3C and table S7**) with smaller degree of chewing movement (ANGLE; electronic supplementary material, **figure S3D and table S7**). This change in cranio-dental morphology between Eocene and Oligocene taxa, which reflects an underlying diet modification, has to be put in perspective with the drastic vegetation change during the Oligocene in Europe. Tropical rainforests of the Eocene were replaced by an open and drier environment during the Oligocene with more abrasive vegetation [88–90] requesting stronger masticatory force/surface for herbivores [42–44]. In addition, our results indicate a significant effect of abiotic factors over the six traits measured, which are associated with body mass and diet (electronic supplementary material, **table S8**), suggesting that environmental changes impacted the cainotheroid trait evolution over the considered time period. The relationship between environment change and species morphology – corresponding to diet – has previously been hypothesised at the artiodactyl scale around the EOT in [28] and is formally tested here. Our results thus highlight the great morphological adaptability of cainotherioid species to major environmental changes of the EOT. This ability to adapt and also, to rapidly diversify, may have played a key role in the survival of the EOT and subsequent evolutionary success of this clade during the early Oligocene.

# Conclusion

The multi-site Konzentrat-Lagerstätte from Quercy (South-West France) documents with unprecedented accuracy the fossil record of the cainotherioid superfamily and allows the analysis of the diversity dynamics of this mammal clade. The estimated preservation rate of ~14 occurrences/taxon/Myrs is comparable to those recovered for invertebrate marine organisms. The diversity dynamics history of cainotherioids is characterized by extinction events at the EOT and at the late Oligocene, and by a major speciation burst in the early Oligocene. Contrary to most other endemic artiodactyl groups, cainotheroids diversify after the EOT thanks to high morphological adaptability, facing environmental changes. Oligocene species display a larger chewing surface and more robust mandibles, fitting the more abrasive vegetation of the drier environmental conditions. The rapid response of cainotheroids to environmental changes likely played a key role in the survival and post-crisis evolutionary success of the clade. We further highlight that speciation is positively associated with temperature fluctuations and continental fragmentation in a time-continuous way, while extinction seems to synchronize with environmental change in a punctuated way. Success and decline of cainotherioids was also partly driven by biotic factors. We find that cainotherioid speciation rates decreased as they diversified over time, and we suggest that within-clade interactions negatively affect the cainotherioid diversification. Finally, we hypothesise that inter-clade competition with Asian newcomers might explain their final decline during the late Oligocene.

**Data accessibility.** Datasets are available from the Dryad Digital Repository: https://doi.org/10.5061/dryad.3tx95x6fn [91]. Additional information are included in the electronic supplementary material.

**Authors' contributions.** R. W., F. C. and M. O. designed the study; R. W. provided data; R. W. and G. G. carried out the statistical analyses; R. W. and F. C. performed data analyses; R. W. and M. O. drafted the manuscript. All authors contributed to the writing, gave final approval for publication and agreed to be held accountable for the work performed therein.

**Competing interests.** We declare we have no competing interests.

**Funding.** This work was financially supported by the ANR program DEADENDER (ANR-18-CE02-0003-01) – P.I. M. J. Orliac.

**Acknowledgements.** We are grateful to T. Pélissié (PNR des Causses du Quercy), the Cloup d'Aural and the Quercy research team (M. Godinot, MNHN, Paris; ISEM, Montpellier; G. Escarguel, LEHNA, Lyon; C. Bousquet, Cloup d'Aural) for their work in the field. We thank S. Jiquel and A.-L. Charruault for her help in the collections of the University of Montpellier. We thank J. Maugoust for his help in statistical treatments with R. We are also grateful to S. Agret for kindly welcoming us in her office to take measurements of the material. Finally, we thank the anonymous reviewers for their constructive comments on an earlier version of the manuscript. This is ISEM publication 2021-061.

# Figure captions

**Figure 1.** Illustration of the evolutionary dynamics of cainotherioid diversity as controlled by time-variable speciation and extinction. With analyses at the species level under the birth–death model with constrained shifts (1 Myrs bins), speciation (A) and extinction (B) rates are inferred. (C) The net diversification rate is the difference between speciation and extinction rates (net diversification rate below 0 indicating a decline in diversity). (D) The cainotherioid diversity trajectories incorporating uncertainties around the age of the fossil occurrences. Solid lines indicate mean posterior rates and the shaded areas show 95% CI. The temperature curve is derived from [53].

**Figure 2.** Diversity dynamics and size evolution (LMRL, body mass proxy) of cainotherioids around the Eocene-Oligocene transition. In light blue, the Robiacinidae family; in blue, the first Cainotheriidae species; and in pink, the Cainotheriinae subfamily. The temperature curve is derived from [53].

**Figure 3.** Morphological evolution of cainotherioids during, before and after the EOT (by MP level) illustrated by (A) diastemata length (DIAS); (B) mandibular corpus height (HCM3). In blue, the first Cainotheriidae species; and in pink, the Cainotheriinae subfamily. The temperature curve is derived from [53]; representations of additional traits are provided in the electronic supplementary material (figure S3).